\documentclass[useAMS,usenatbib]{mn2e}
\usepackage{graphicx}
\usepackage{times}

\title[Turbulence  at Cerro Tololo]
{Statistics of turbulence profile at Cerro Tololo}

\author[A.~Tokovinin, S.~Baumont, J.~Vasquez]
{A.~Tokovinin\thanks{E-mail:
atokovinin@ctio.naoo.edu},
S.~Baumont, J.~Vasquez \\
Cerro Tololo Inter-American Observatory, Casilla 603, La Serena, Chile}

\begin{document}

\date{Accepted 2002 Month dd. Received }
\pagerange{\pageref{firstpage}--\pageref{lastpage}} \pubyear{2002}

\maketitle

\label{firstpage}

\begin{abstract}

Results  of 3-month  continuous monitoring  of turbulence  profile and
seeing at Cerro  Tololo (Chile) in May-July 2002  are presented.  Some
28000 low-resolution profiles were  measured by a new MASS single-star
turbulence monitor, accompanied by  seeing data from DIMM.  The median
seeing  was 0.95  arcseconds. The  first 500~m  contribute 60\%  to the
total seeing,  the free-atmosphere median seeing  was 0.55 arcseconds.
Free-atmosphere  seeing is  almost never  better than  0.15 arcseconds
because there is always some turbulence above 12~km. A 4-day period of
calm upper atmosphere with  a stable free-atmosphere seeing of 0.2-0.3
arcseconds was noted.  A gain in resolution from adaptive compensation
of ground layer will be 1.7  times typically and 2-3 times during such
calm periods.  Correlations of the free-atmosphere turbulence with the
wind  speed at  tropopause  and of  the  ground-layer turbulence  with
ground  wind  are  studied.    Temporal  evolution  of  turbulence  is
characterized  by recurrent bursts,  their typical  duration increases
from 15 minutes  in low layers to 1-2 hours in  high layers. The large
data  base of  turbulence  profiles  can be  used  to test  meso-scale
modeling of astronomical seeing.
\end{abstract}


\section{Introduction}

A crucial role of  `seeing' in ground-based astronomy was recognized
long time  ago.  Nowadays  it is possible  to improve the  seeing with
adaptive optics  (AO), but this  technology is itself so  dependent on
the properties  of turbulence  that it generated  a new  and important
driver for  detailed atmospheric studies.  AO requires a  knowledge of
the temporal time constant and  of the vertical turbulence profile, in
addition to the overall (integrated) seeing. It is desirable to have a
reliable statistics of  these parameters for a given  site in order to
predict  the performance  of AO  systems.  A  real-time  monitoring of
optical  turbulence would help  in optimizing  the AO  operation.  For
example, the  scintillometer of Ochs.  et al.   was regularly operated
at the AMOS station for this reason \citep{Chonacky88}.

The   vertical  distribution  of   the  optical   turbulence  strength
(characterized  by the  altitude  dependence of  the refractive  index
structure    constant   $C_n^2$)    is   notoriously    difficult   to
monitor. Balloon-born micro-thermal probes  are expensive and sample the
turbulence profile (TP) only  once per flight, without any statistical
averaging.   Optical remote sounding  by SCIDAR \citep{Fuchs98}  is free
from  this  drawback, but  it  requires  moderately large  telescopes,
sensitive detectors, and powerful  signal processing. For these reasons
SCIDAR was only used in  a campaign mode at existing observatories.

A limited  number of TPs measured world-wide  revealed that turbulence
is typically concentrated in  few thin layers.  The physical mechanism
generating  such  distribution  was  studied by  \citet{Coulman}.   It
inspired designers of AO systems  to add more deformable mirrors, each
conjugated to its own layer, and thus to compensate seeing over a much
wider  field  with  such  multi-conjugate AO  (MCAO).   The  promising
potential  and wide  popularity of  MCAO added  even more  pressure to
measure turbulence profiles; the Gemini site testing campaign at Cerro
Pach\'on   \citep{Vernin2000,Avila2000}   is   an  example   of   such
MCAO-driven study.

Ground-based telescopes of next  generation with apertures of 20-100~m
will include  turbulence compensation already in  their designs. Sites
for these telescopes are being  selected with a strong weight given to
AO-related  turbulence  parameters;   site  surveys  based  on  seeing
measurements alone,  as was  the case for  the previous  generation of
telescopes, are  no longer sufficient.  Seeing is  very much dominated
by local and orographic effects  that diminish predictive power of
seeing  data.  With  modern  computers,  a  modeling of  optical
turbulence becomes  feasible, giving new insights into  the physics of
seeing    and    new    guidance    to    the    choice    of    sites,
e.g. \citep{Masciadri2001}.  But computer  models still need real TPs
for their calibration.

A   low-resolution    turbulence   profile   monitor,   Multi-Aperture
Scintillation Sensor (MASS), was developed in response to the needs of
AO  and  MCAO, as  well  as a  portable  instrument  for site  testing
\citep{Marrakesh,MASS}.  MASS  was operated in 2002  for several months
at the Cerro Tololo Inter-American Observatory (CTIO) jointly with the
Differential Image Motion  Monitor (DIMM) \citep{CTIODIMM}.  This paper
presents  the results  of this  campaign. It  appears to  be  the most
extensive data base of turbulence profiles existing to date worldwide.

Our aim was to gain  some understanding of the turbulence localization
above  CTIO.   We were  specifically  interested  in  the fraction  of
turbulence in the first few hundred  meters over the ground and in the
seeing that  can be attained if  these low layers  were compensated by
AO. {\em  Ground-layer compensation} offers improved seeing  in a much
wider field than does classical AO \citep{Rigaut2001}.  This option is
being studied  for the 4.2-m  SOAR telescope located close  to CTIO on
Cerro Pach\'on \citep{SOAR}, as well as for extremely large telescopes
of  next  generation.  Our  work  quantifies  the  gain expected  from
ground-layer compensation at a specific good astronomical site, CTIO.
 
In Sect.~2 we briefly describe  the instrumentation used in this study
and give typical examples of the data.  The statistics of the vertical
turbulence  distribution  is explored  in  Sect.~3.  Sect.~4  contains
summary and conclusions.

\section{Site, instruments and data}

\subsection{Site}

\begin{figure}
\includegraphics[width=7cm]{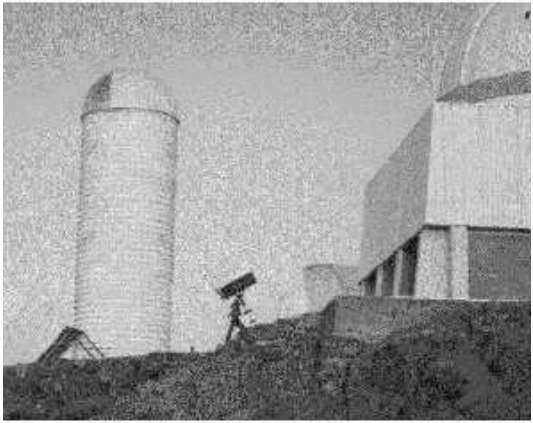} 
\caption{\label{fig:location}  
The DIMM instrument  is installed  in a
6~m high tower at the edge  of Cerro Tololo summit. MASS was initially
located on the  ground, as shown; it was moved into  the USNO dome (on
the right) on May 19 2002. 
 } 
\end{figure}

Cerro Tololo is located in Chile  some 500~km to the north of Santiago
($70\degr 48\arcmin 52\farcs 7$ W, $30\degr 09\arcmin 55\farcs 5$ S). Altitude
is 2200~m above  sea level (a.s.l.). Among other  Chilean sites, Cerro
Tololo is known for its low ground wind speed.

The   two   instruments   used   in  this   campaign,   Multi-Aperture
Scintillation  Sensor  (MASS) and  Differential  Image Motion  Monitor
(DIMM), were  placed at  the northern edge  of the summit,  facing the
direction  of  prevailing  wind  (Fig.~\ref{fig:location}).   DIMM  is
placed  in a  small tower  at  some 6~m  above the  ground.  The  MASS
feeding  telescope was  installed  on the  Losmandy equatorial  mount,
initially on the ground and later in a small dome.


\subsection{MASS}

MASS   \citep{Marrakesh,MASS}   measures   low-resolution   turbulence
profiles  from  the  scintillation  of  single stars.  Light  flux  is
received by  four concentric-ring apertures with diameters  of 2, 3.7,
7.0, and  13~cm and detected  by photo-multipliers in  photon counting
mode with 1~ms time sampling.  Statistical analysis of the fluxes with
1~minute  accumulation  time  produces  10 scintillation  indices  that
correspond  to  4  individual   apertures  and  6  pair-wise  aperture
combinations.  MASS  is fed by  a 14-cm off-axis  reflecting telescope
specially designed  for this purpose.  

At the  beginning of a  night, MASS is  pointed to a  bright ($V<2^m$)
blue single  star close to  zenith.  After background  measurement and
star  centering,  a  series  of continuous  1-minute  integrations  is
started,  with either manual  or automatic  guiding of  the telescope.
When  zenith distance  of  the star  increases  above $45^\circ$,  the
telescope is  re-pointed to another object  (a total of  3-4 stars per
night).

\begin{figure}
\includegraphics[width=7cm]{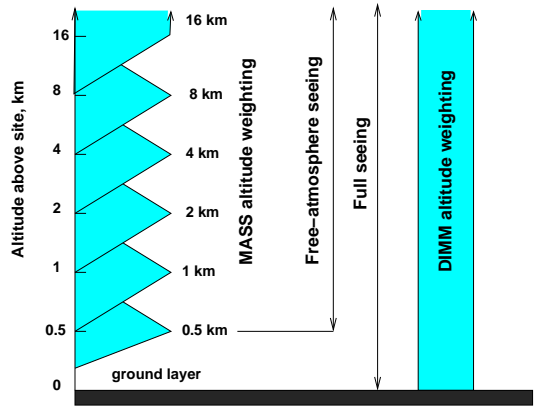} 
\caption{\label{fig:layers} Scheme of  the sub-division of atmospheric
path into 0.5-km, 1-km, etc. layers  measured by MASS and the total path
measured by DIMM. Turbulence in  the ground layer is inferred from the
difference between DIMM and MASS.  }
\end{figure}

A model of turbulence distribution with 6 layers at fixed altitudes of
0.5, 1, 2,  4, 8, and 16~km  above ground is fitted to  the data. Each
`layer' represents in fact an  integral of turbulence $J_l$ 

\begin{equation}
J_l = \int_{\rm layer}  C_n^2(h) W(h) {\rm d}h
\end{equation}
measured  in  m$^{1/3}$,  where  $C_n^2(h)$  is  the  refractive-index
structure  constant  in  m$^{-2/3}$  and $W(h)$  is  the  dimensionless
response  function of  the  instrument.  MASS  response functions  are
nearly triangular, going  to zero at the altitudes  of adjacent layers
(Fig.~\ref{fig:layers}).   Thus,  the  8-km layer  measures,  roughly,
integrated  turbulence strength  from 6  to 12~km,  while  16-km layer
measures everything  above 12~km. Turbulence  at 6~km will show  up in
4-km and 8-km `layers' with  equal intensity.  The sum of all response
functions is constant (within 10\%) at altitudes above 0.5~km.

Atmospheric  seeing $\epsilon_l$  (full  width at  half  maximum of  a
long-exposure  image in  a large  telescope) produced  by  a turbulent
layer can  be computed from  the intensity $J_l$. For  $\epsilon_l$ in
arcseconds at $\lambda =500$~nm and $J_l$ in m$^{1/3}$,

\begin{equation}
\epsilon_l = (J_l / 6.8 \cdot 10^{13})^{3/5}.
\end{equation}

The  sum  of all  layer  intensities measured  by  MASS  gives a  good
estimate of  the `free-atmosphere  seeing' $\epsilon_f$ --  the seeing
that would be  obtained without contribution of the  turbulence in the
first  500~m  above  ground.    The  free-atmosphere  seeing  is  also
estimated  by MASS  directly  from a  combination  of 3  scintillation
indices.  There is a very good agreement between $\epsilon_f$ computed
from  the  integrated profiles  and  directly.   More  details on  the
restoration procedure are given in \citep{MASSrestor}.  It is shown in
this study  that noise in  MASS is signal-dependent,  becoming smaller
under low-turbulence conditions.  The strength of dominating layers is
always measured with a typical relative  error of 10\% which may be as
low as $4 \cdot 10^{-15}$ m$^{1/3}$ under calm conditions.

MASS data  are reduced  in real time  and stored  in an ASCII  file, a
profile  every  minute.    Many  additional  parameters  (instrumental
configuration,  stellar fluxes,  quality of  the model  fit)  are also
written to  this file. This permits  to identify and  reject any wrong
data.   The  reasons for  erroneous  data  are  guiding errors,  wrong
background estimates, and, by far most numerous, clouds.  It was found
that  MASS  gives consistent  and  reproducible  results through  thin
cirrus clouds because slow  light variations (below 1~Hz) are filtered
out in the  data reduction algorithm.  We rejected  only the data with
low stellar fluxes and with flux  variance (in 1 ~minute with 1~s flux
averaging) of more than  1\% that indicated varying cloud transmission
during integration.

\subsection{DIMM}

DIMM measures  the Fried  parameter $r_0$ related  to the  full seeing
$\epsilon   =    0.98   \lambda   /r_0$    at   wavelength   $\lambda$
\citep{DIMM,Tok2002}.  The seeing is  deduced from the variance of the
angle-of-arrival   fluctuations  (or  image   motion)  is   two  small
apertures.   

The CTIO DIMM \citep{CTIODIMM}  uses 25~cm Meade as feeding telescope.
The  diameter of the  apertures is  95~mm (partially  obstructed), the
distance between  their centres is  15.3~cm.  Images of a  bright star
formed by both apertures are separated  by a wedge prism placed on one
entrance  aperture and detected  by a  CCD camera  with pixel  size of
$0\farcs 77$.  Frame exposure time  is alternating between 5~ms and 10~ms,
the integration time for seeing estimate is 1~minute, with acquisition
rate about  300 images per  minute.  Upon background  subtraction, the
centroids of images  are computed in a window  of 8~pixel radius.  The
variance  of   centroid  coordinate  differences   in  two  orthogonal
directions is  corrected for noise  variance and converted  to seeing.
The bias in seeing caused by  finite exposure time is corrected by the
modified exponential prescription as detailed in \citep{Tok2002}.  The
DIMM data are  stored in an ASCII file and  are also available through
WEB in real time.

DIMM  operates  in a  robotic  mode,  opening  its dome  and  pointing
suitable stars when the meteo conditions are adequate. Guiding is done
in-between  1-minute   data  accumulations.   Robotic   operation  was
interrupted repeatedly  by failures  to find a  star which  called for
manual interventions (the Meade  mount does not have absolute position
sensors to recover its pointing)  and by failures of the drive motors,
replaced several times.  For this  reason the time coverage of DIMM is
somewhat less than that of MASS for the same period.

\subsection{Comparison between MASS and DIMM}

\begin{figure}
\includegraphics[height=6cm]{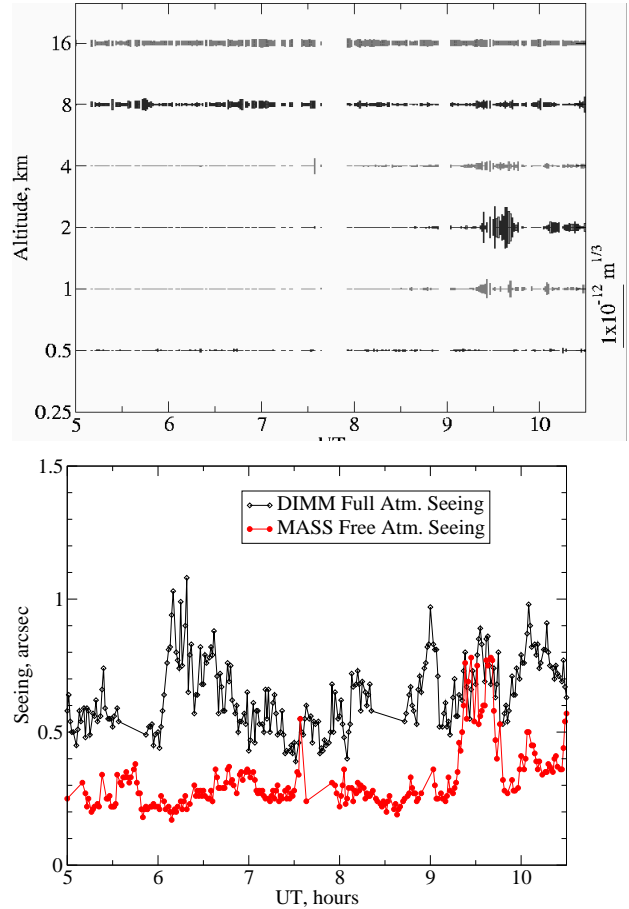} 
\includegraphics[height=6cm]{020619fsee.eps}  
\caption{\label{fig:jun19}  Turbulence   profiles  (top)  and  seeing
(bottom) on a very calm night of June 19-20, 2002. Bars  show the
intensity of layers in m$^{1/3}$  with a scale indicated on the right.
}
\end{figure}

\begin{figure}
\includegraphics[width=8.4cm]{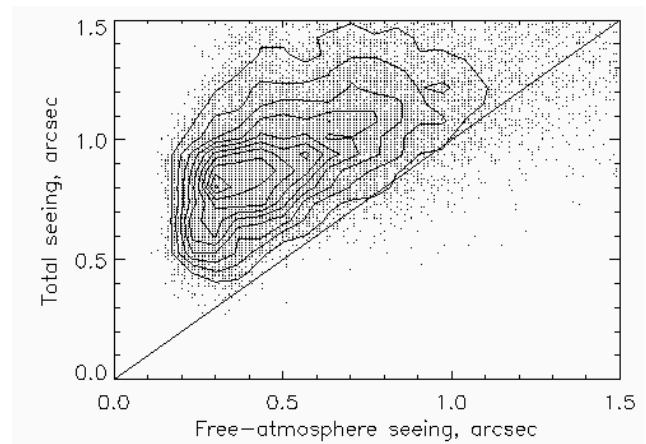} 
\caption{\label{fig:dimm-mass}   Comparison  of   the  free-atmosphere
seeing $\epsilon_f$  measured by MASS  (horizontal axis) and  the full
seeing $\epsilon$  measured by  DIMM (vertical axis)  for all  data in
common.   Contour  plots  are  over-layed  with  10\%
intervals to show the density of points.  The line shows $\epsilon_f =
\epsilon$.}
\end{figure}

MASS  and   DIMM  measure  different   `seeings',  $\epsilon_f$  and
$\epsilon$. Whenever the contribution  of the first 500~m above ground
to  the total  seeing is  small,  we must  obtain $\epsilon_f  \approx
\epsilon$, otherwise the inequality $\epsilon_f < \epsilon$ must hold.
This is indeed the case.  In Fig.~\ref{fig:jun19} we display a portion
of data  for the night of  June 19-20 2002 with  very calm atmospheric
conditions.  Around  9.5h UT an  increased turbulence at  2~km dominated
the seeing, with both instruments giving very similar results. For the
rest of the night, the full seeing $\epsilon$ was very good, but still
worse than $\epsilon_f$ because of ground-layer turbulence.

The condition $\epsilon_f \approx  \epsilon$ illustrated above was not
exceptional,    but,     on    the    contrary,     was    encountered
regularly. Dominating  high layers were  frequently located at  0.5 --
1~km, but  could be as  high as 8~km.   The agreement between  the two
very different  instruments, MASS and  DIMM, is better than  one could
expect; it gives confidence that both produce correct data.

Fig.~\ref{fig:dimm-mass} shows free-atmosphere  seeing compared to the
full seeing  measured simultaneously  (interpolated to the  moments of
MASS measurement  from two  adjacent DIMM points)  for the  whole data
set. The occasional equality  $\epsilon_f \approx \epsilon$ is clearly
seen.   A  small number  of  points  is  located below  the  diagonal,
$\epsilon_f > \epsilon$. These points can be explained by noise and by
the fact that both instruments used different stars, so small localized
patches  of  turbulence occasionally  caused  spikes  in the  MASS
seeing not matched by DIMM. The inverse is also happening, of course.

\subsection{Data overview}

\begin{figure}
\includegraphics[width=8.5cm]{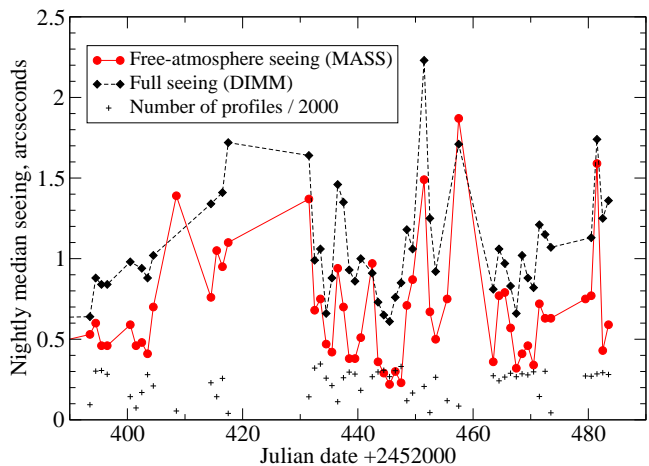} 
\caption{\label{fig:coverage}  Nightly median  values of  total seeing
$\epsilon$  and  free-atmosphere seeing  $\epsilon_f$  for the  period
April 29  to July  28 2002.  The  crosses show  a total number  of TPs
acquired each night (less than 500 on partially cloudy nights). }
\end{figure}

First useful  MASS data were obtained  on March 21-27  2002 during the
commissioning  of  this  instrument.   Systematic  profile  monitoring
started on  April 29. Here  we consider the  data obtained to  July 28
inclusive, with  3 complete months  covered.  During this  period MASS
worked for  58 nights, but some  of them were partially  cloudy with a
reduced number  of data. On a  clear night MASS  measured some 500-600
profiles.  The  total number of TPs  used for the  analysis (after the
cleaning mentioned above)  is 22300, of which 16968  TPs have matching
DIMM data.

The  periods of  bad weather  are apparent  in Fig.~\ref{fig:coverage}
where  nightly  median  values  of  $\epsilon$  and  $\epsilon_f$  are
plotted. A particularly long cloudy period occurred between May 11 and
June 6.

All seeing  and TP  data in  this article refer  to the  wavelength of
500~nm and to zenith.  The following results characterize atmosphere
at  Cerro Tololo  in the  period May-July  2002.  This  corresponds to
autumn and winter conditions, typically worse than average.

\section{Data analysis}

\subsection{Relative contribution of atmospheric layers to seeing}

\begin{table}
  \caption{\label{tab:seeing}  Levels of the  cumulative distributions
  of  total seeing  $\epsilon$,  free-atmosphere seeing  $\epsilon_f$,
  ground-layer seeing  $\epsilon_g$, fraction of ground layer, 
   and isoplanatic  angle $\theta_0$. }
  \begin{tabular}{ l | ccccc }
   \hline
    Probability    & 10{\%} & 25{\%} & 50{\%} & 75{\%} & 90{\%}\\
   \hline
    Total seeing $\epsilon$, arcsec        &0.64 &0.79 &0.95 &1.17 &1.43\\
    Free-atm. seeing $\epsilon_f$, arcsec  &0.28 &0.38 &0.55 &0.82 &1.22\\
    Ground-layer seeing $\epsilon_g$, arcsec &0.24 &0.47 &0.66 &0.83 &1.02\\
    Ground-layer fraction $f_g$    &   0.11 & 0.38 & 0.60 & 0.76 & 0.85 \\    
   Isoplanatic angle $\theta_0$, arcsec & 2.94 &2.36 &1.80 &1.30 &0.98\\ 
  \hline
  \end{tabular}
\end{table}

\begin{figure}
\includegraphics[width=8.5cm]{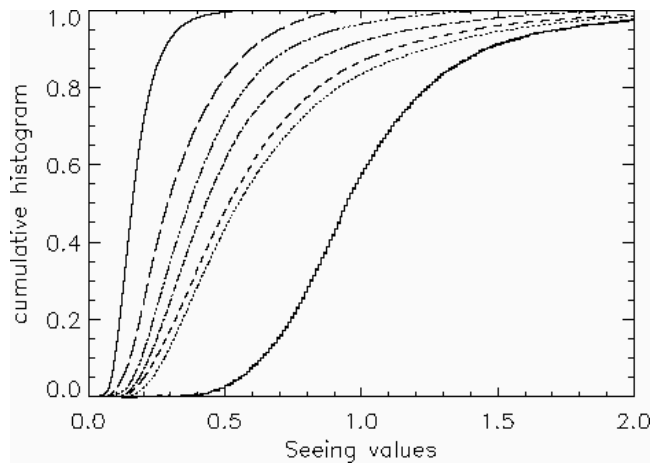} 
\caption{\label{fig:hist} Cumulative distribution  of the total seeing
(thick line) and (from right to  left) of the seeing that would result
from correction of the ground layer, first 1~km, first 2~km, etc. 
to the seeing produced by 16~km layer alone (leftmost curve).
}
\end{figure}

\begin{table}
  \caption{\label{tab:corr}  Levels of  the cumulative  distributions of
seeing  in arcseconds  for  different thickness  of corrected  layers.}
 \begin{tabular}{ l | ccccc }
 \hline
   Probability       & 10{\%} & 25{\%} & 50{\%} & 75{\%} & 90{\%}\\
 \hline
   No correction ($\epsilon$)    &0.64 &0.78 &0.95 &1.17 &1.46\\
   Ground to 0.5 km ($\epsilon_f$) &0.28 &0.38 &0.55 &0.83 &1.24\\
   Ground to 2 km    &0.22 &0.30 &0.43 &0.62 &0.92\\
   Ground to 4 km    &0.19 &0.26 &0.37 &0.53 &0.73\\
   Ground to 8 km    &0.15 &0.20 &0.29 &0.43 &0.60\\
   Ground to 16 km   &0.10 &0.12 &0.16 &0.21 &0.28\\
 \hline
 \end{tabular}
\end{table}

Table~\ref{tab:seeing}   contains  the   main  levels   of  cumulative
distributions    of   seeing:   total    $\epsilon$,   free-atmosphere
$\epsilon_f$, and ground-layer  seeing $\epsilon_g = (\epsilon^{5/3} -
\epsilon_f^{5/3})^{3/5}$. In 7\% of cases with $\epsilon_f > \epsilon$
we assume that  ground layer seeing was close to  zero. The data used
for  this  analysis  refer only  to  cases  when  both MASS  and  DIMM
measurements are  available. We also  include the distribution  of the
isoplanatic angle $\theta_0$ which is readily computed from MASS data.

Supposing that  ground-layer turbulence can be  measured and corrected
by AO, we can estimate  the resulting improvement in seeing. Of course,
realistic  adaptive optics  will not  correct  ground-layer turbulence
perfectly, but,  on the other  hand, it will partially  correct higher
layers. Still,  the analysis presented in  Table~\ref{tab:corr} and in
Fig.~\ref{fig:hist} gives  a quantitative idea on the  gain in angular
resolution expected  from ground-layer correction.  The gain increases
when a thicker  slab of turbulence is corrected, but  at the same time
the size of corrected field becomes smaller.

Table~\ref{tab:seeing} also contains  the distribution of the fraction
of $C_n^2$  integral contained  in the first  500~m above  the ground,
$f_g  = (\epsilon_g  /\epsilon  )^{5/3}$.  We  studied this  parameter
separately for good ($<0\farcs 6$)  and bad ($>1 \arcsec $) seeing and
did  not  find  any   substantial  dependence.   For  example,  median
fractions of ground  layer for good and bad seeing  are 0.57 and 0.53,
respectively.  The largest median contribution of ground layer -- 0.66
-- is found for seeing between $0\farcs 6$ and $1 \arcsec$. Turbulence
profiles   measured  at   the  nearby   mountain  Cerro   Pach\'on  by
\citet{Vernin2000} led to a model where average contribution of ground
layer is 65\% \citep{Ellerbroek2000}.

Our  results  should  be  compared  to  the  extensive  monitoring  of
turbulence profile at Cerro Tololo with Generalized SCIDAR reported by
\citet{Vernin2000} and \citet{Avila2000}.  These authors obtained 6900
TPs over  22 nights distributed  in 4 campaigns throughout  year 1998.
They find  median total seeing of  $\epsilon = 1  \farcs 06$ ($0\farcs
85$ when dome contribution is excluded), median free-atmosphere seeing
$\epsilon_f  = 0\farcs 56$  and median  isoplanatic angle  $\theta_0 =
2\farcs  14$.   The agreement  with  our  data  is encouraging.   Free
atmosphere was  more perturbed in  April and July compared  to January
and October  1998.  Thus,  we expect that  somewhat better  seeing and
larger isoplanatic angles will be measured during summer months.

\citet{Barletti76} proposed a TP model  based on 67 micro-thermal
soundings. It  predicts an average free-atmosphere  seeing (all layers
above 1~km)  of $0\farcs 80$.  Their `lucky observer'  model assumes the
lowest  turbulence  levels  measured  at each  altitude  and  predicts
$\epsilon_f =  0\farcs 28$. We  see from Table~\ref{tab:seeing}  that such
conditions are  indeed encountered  at Cerro Tololo  10\% of  time and
that   the   median   $\epsilon_f$   is  significantly   better   than
$0\farcs 80$. \citet{Marks2002}  used 15 soundings in  Antarctica to claim
that  upper  atmosphere  there  is  exceptionally calm,  with  a  mean
$\epsilon_f  = 0\farcs 37$. Equivalent  or better  conditions do  occur at
Cerro Tololo 25\% of time even during unfavorable winter months.

It is  interesting that  even when all  layers except the  highest are
corrected,  the expected  seeing  is almost  {\em  never} better  than
$0\farcs  15$.    The  same  conclusion  is   strikingly  apparent  in
Fig.~\ref{fig:dimm-mass} where  $\epsilon_f$ has a  sharp lower cutoff
at $0\farcs 15$. There is always some non-negligible turbulence at the
upper  boundary  of the  troposphere  (see also  Fig.~\ref{fig:wind}).
Median $C_n^2$  integral in  the 16~km layer  is $3.2  \cdot 10^{-14}$
m$^{1/3}$, some 10  times higher than the instrumental  noise of MASS,
which  means  that  the  effect  is  real  and  not  related  to  some
instrumental threshold.  \citet{Vernin2000} measured the median seeing
produced between 15~km and 20~km as $0\farcs 14$.

The consequences  of this finding  are important for  adaptive optics.
Even when the turbulence profile is completely dominated by few strong
and sharp layers,  their correction by means of  MCAO will not suffice
to reach diffraction-limited resolution in a wide field because of the
remaining high  layers.  The numbers  in Table~\ref{tab:corr} indicate
median  fraction of  the 16~km  layer as  5\%; a  similar  fraction of
high-altitude  turbulence is  adopted in  MCAO simulations  for Gemini
\citep{Ellerbroek2000}.   Another consequence of  the upper-troposphere
layers is  the effective limit  on isoplanatic field  size $\theta_0$:
even  under very calm  conditions it  practically never  exceeds $\sim
5\arcsec$ as measured at several sites. Median turbulence measured at 16~km
layer  alone would give  isoplanatic angle  $\theta_0 =  13\farcs 4$.

\subsection{Temporal variation of turbulence}

\begin{figure}
\includegraphics[width=8cm]{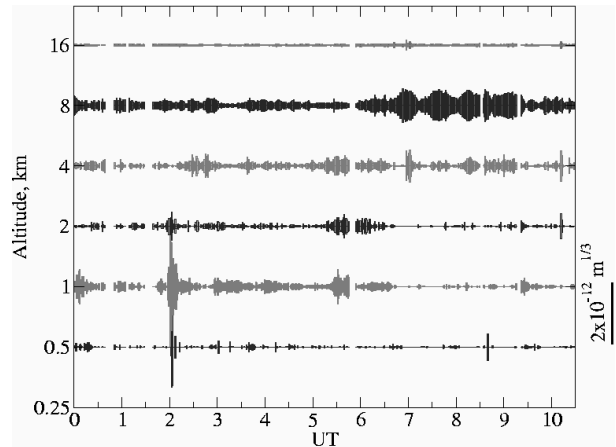} 
\caption{\label{fig:Jul25} A typical night (July 25-26 2002) with
 turbulence bursts. 
  }
\end{figure}

\begin{figure}
\includegraphics[width=8cm]{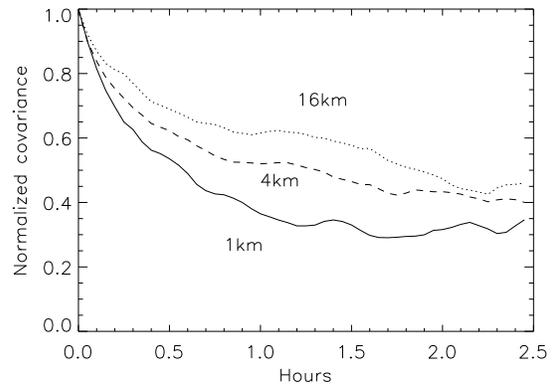} 
\caption{\label{fig:temp}  
Normalized temporal autocorrelation
functions of turbulence strength in three representative layers for
the whole TP database.
  }
\end{figure}

Strong bursts of turbulence that last typically for less than 0.5 hour
were observed repeatedly in almost all layers. An example of a typical
night   with   perturbed   atmosphere   and   bursts   is   given   in
Fig.~\ref{fig:Jul25}  (note  the coarser  vertical  scale compared  to
Fig.~\ref{fig:jun19}).

In Fig.~\ref{fig:temp}  the normalized temporal autocorrelation
functions  $A(\Delta t)$  for low,  intermediate and  high  layers are
plotted. The data -- layer intensities $J(t)$ -- are not evenly spaced
in time and have some  gaps, making computation of covariance somewhat
difficult.  For  this purpose we  re-binned the $J(t_i)$ on  a uniform
grid  with  3~minute step,  padding  missing  data  with zeroes.   The
covariance  $C(\Delta  t)$  was  then computed  and  transformed  into
autocorrelation $A(\Delta t)$:

\begin{eqnarray}
C(\Delta t) & = & \frac{1}{N} \sum J(t_i) J(t_i + \Delta t), \\
A(\Delta t) & = & \frac{C(\Delta t) - \overline{J}^2 }
{C(0) - \overline{J}^2 },
\end{eqnarray} 
where $N$  is the  total number  of non-zero products  for a  time lag
$\Delta t$ and $ \overline{J}$~ is the average of $J$.

It  can be  seen that  the time  constant of  turbulence  variation is
longer  for  high  layers  and   shorter  for  low  layers.   At  50\%
correlation this  time is 0.6, 1.3 and  1.9 hours for 1,  4, and 16~km
layers  respectively.   The  autocorrelation  functions  show  a  fast
decrease  at small time  lags that  correspond to  a presence  of more
rapid variations  and are  suggestive of power-law  temporal spectrum.
On the other  hand, the `tails' of $A(\Delta  t)$ reflect long-term,
night-to-night  variations.  If  we subtract  the tail  from  the 1~km
autocorrelation,  the 50\%  correlation time  would become  0.25 hour,
giving a better  idea on the duration of  turbulent bursts. Bursts are
shorter than  periods between them,  but $A(\Delta t)$  cannot reflect
this difference.

Rich  data on  TPs enable  a better  understanding of  the  physics of
atmospheric  seeing.   Seeing  variations  usually  observed  at  good
astronomical   sites  with   a  time   scale  of   $\sim   1.2$~h
\citep{Sarazin97}  can  now be  traced  to  the appearance  turbulence
bursts at  specific altitudes.  These  bursts are a  local phenomenon.
We frequently compared  the seeing at Cerro Tololo  with the seeing at
La  Silla, only $\sim$~100~km  away: there  is no  correlation between
seeing variations at these sites.  Turbulence at medium altitudes must
be related  to the orographic  disturbances.  If at some  lucky summit
this turbulence  could be  avoided, it would  enjoy a better  and more
stable  seeing.    Hydrodynamical  modeling  of   turbulent  flows  is
necessary to  understand whether such `lucky  summits' indeed exist.
On the  other hand, it becomes  evident that seeing  statistics at any
specific site is not  necessarily representative of other mountains in
the same region.

\subsection{Seeing and weather}

\begin{table}
  \caption{\label{tab:calm} Nightly  median values of  the full seeing
 $\epsilon$, free-atmosphere seeing $\epsilon_f$, wind speed at 200mb,
 ground wind  speed and direction for  a calm period in  June 2002 and
 adjacent dates.  }

 \begin{tabular}{ l | cc cc }
 \hline
   Date,  & $\epsilon$, & $\epsilon_f$, & $V_{\rm 200mb}$, & $V_{\rm ground}$, \\
  2002         & arcsec     &   arcsec     & m/s  &  m/s  \\
 \hline
June 18  & 0.79 & 0.36 & 21.3  & 3.4 S   \\
June 19  & 0.64 & 0.29 & 22.7  & 2.3 S  \\
June 20  & 0.56 & 0.21 & 26.1  & 1.5 S  \\
June 21  & 0.82 & 0.28 & 29.5  & 1.8 E  \\
June 22  & 0.86 & 0.25 & 27.7  & 4.6 E  \\
June 23  & 1.09 & 0.87 & 29.3  & 4.7 E  \\
 \hline
 \end{tabular}
\end{table}

\begin{figure}
\includegraphics[width=8cm]{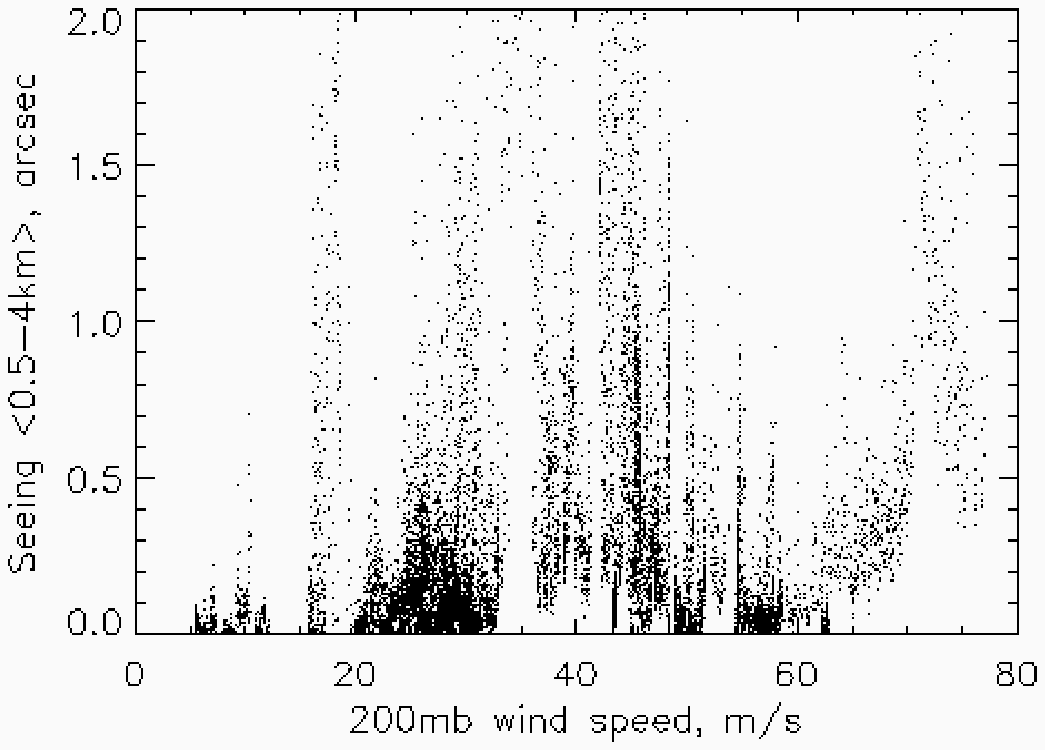} 
\includegraphics[width=8cm]{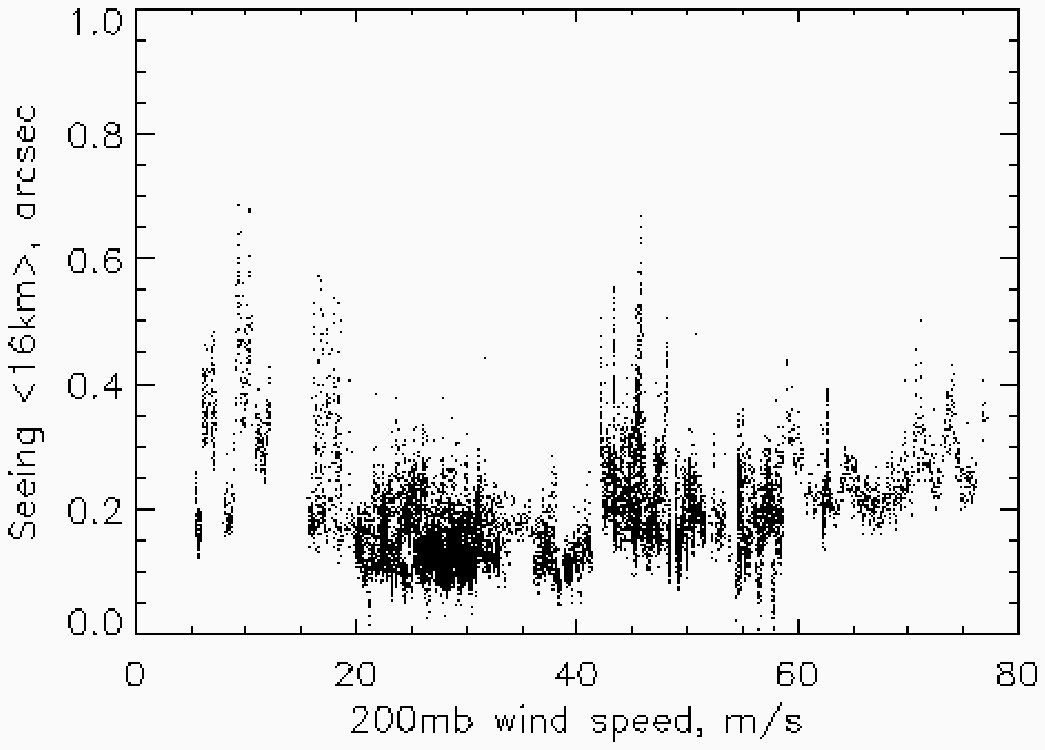} 
\caption{\label{fig:wind}  Correlation   of  the  seeing  
generated by the low atmospheric layers (0.5--4~km, top) and by the 16-km layer
(upper  troposphere, bottom)  with the  wind  speed  at  200~mb level  (12  km
a.s.l.).  }
\end{figure}

\begin{figure}
\includegraphics[width=8cm]{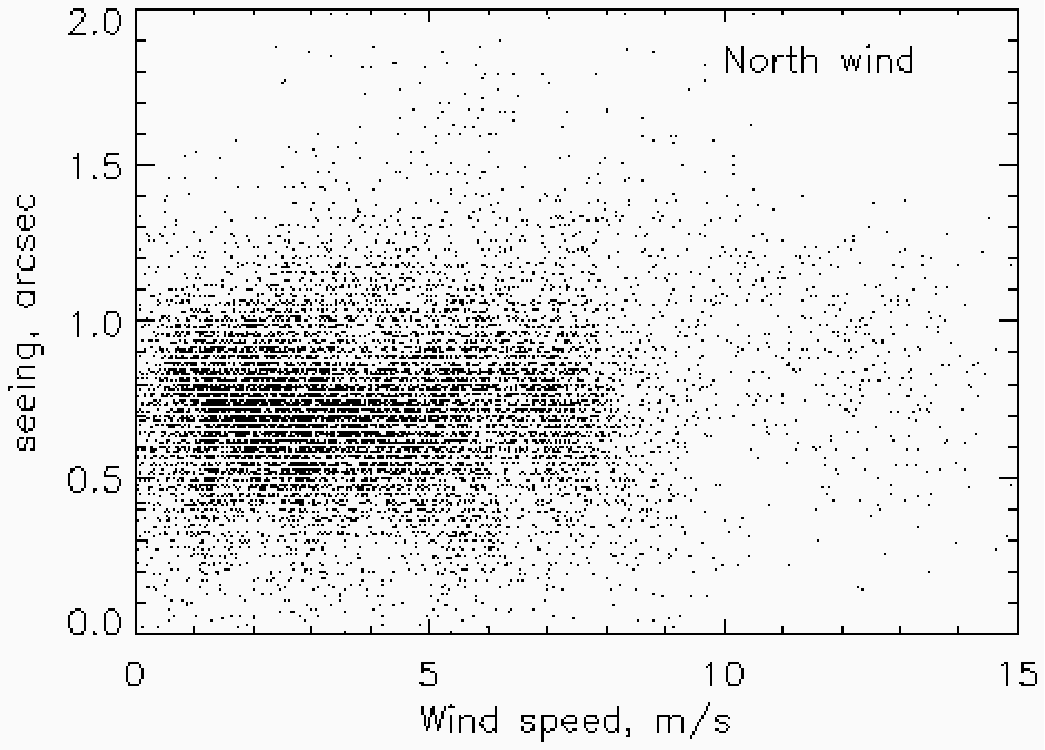}
\includegraphics[width=8cm]{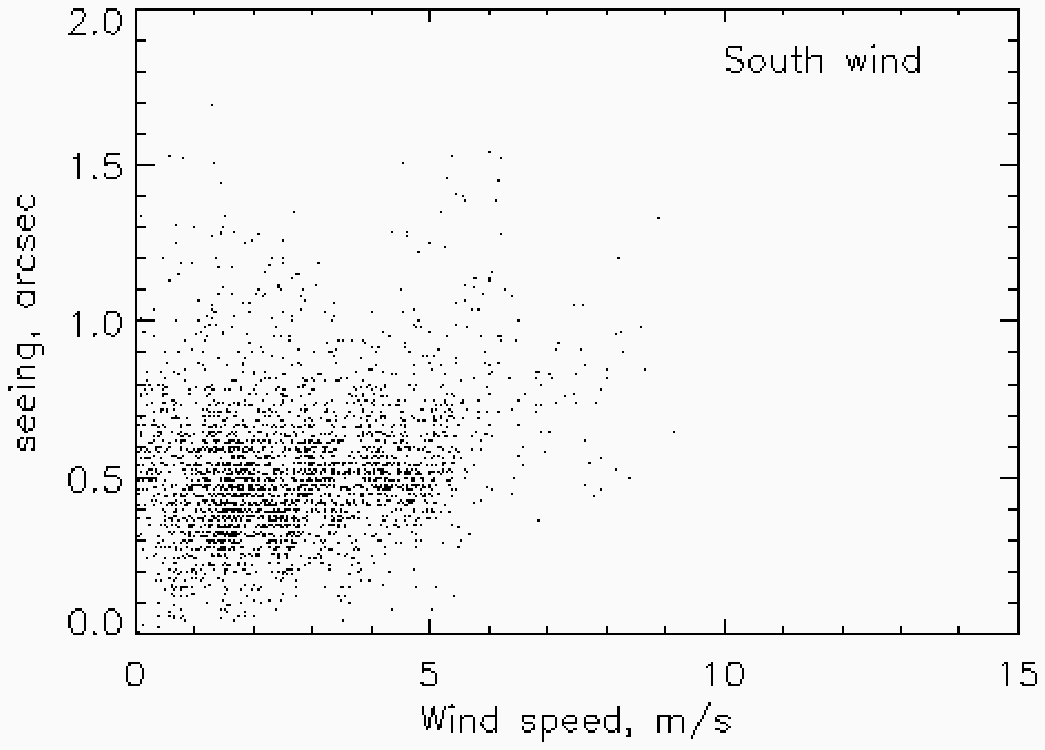} 
\caption{\label{fig:gwind} Correlation  of the seeing $\epsilon_g$ 
generated by the
ground layer (below 0.5~km) with the speed of ground wind plotted
separately for the north-eastern (top) and south-western (bottom) dominating wind
directions.  }
\end{figure}

A   period   of  calm   atmosphere   over   Cerro   Tololo  shown   in
Fig.~\ref{fig:jun19} actually lasted for four consecutive nights, from
June  19 to June  22. The  free-atmosphere seeing  was very  stable at
$0\farcs 2 -  0\farcs 25$; the burst shown  in Fig.~\ref{fig:jun19} is
the worst  $\epsilon_f$ measured during  this whole period.  The total
seeing as measured by DIMM was good but not exceptional. Seeing at the
ESO sites La Silla and Paranal was also only moderately good. Based on
DIMMs alone, one would never tell that something special was happening
in the atmosphere on these  dates. We have continuous profile data for
the period from June 6 to June  30, with only small gaps due to cirrus
clouds,  that bracket this  special period.  Nightly median  values of
seeing, wind in the high atmosphere  and at the ground for this period
are listed in Table~\ref{tab:calm}.

Investigation  of  meteorological  conditions  that  produce  extended
periods of calm atmosphere is  of evident practical interest if it can
lead to understanding and prediction of such periods.  A similar 3-day
calm period was detected by \citet{Avila2002} in May 2000 at San Pedro
M\'artir in Mexico, with  stable values of $\epsilon_f \approx 0\farcs
2$.  This proves  that calm atmosphere is something  not very unusual.
If all  layers were  independent of each  other, a probability  of all
being quiet simultaneously would be low, and a probability of extended
quiet periods  would be vanishingly  small.  Instead, there must  be a
factor  common to  all  layers  that produces  calm  conditions in  a
systematic way, as noted for the first time by \citet{Barletti76}.

As a first  attempt to understand quiet periods,  we analyzed the data
on the speed  of jet stream (wind at 200 mb  pressure or 12~km a.s.l.)
over La Silla  as collected by ESO. Indeed,  the wind velocity $V_{\rm
200mb}$ was low.   We have data on $V_{\rm 200mb}$  only up to July~4,
2002. Seeing produced  by low and  high layers is compared  to $V_{\rm
200mb}$ for this period in Fig.~\ref{fig:wind}.  The period studied is
certainly  not   long  enough   to  cover  all   representative  meteo
conditions; the discussion that follows  may be affected by a specific
combination of weather patterns that occurred during this period.

Looking at  Fig.~\ref{fig:wind}, we note  that there is  no one-to-one
correlation between jet stream  and turbulence, but some tendencies do
emerge. Low layers are generally  more quiet when the jet stream speed
is low. Of course, this may be related to the fact that wind speed and
stability in all layers correlate  with the 200 mb wind. The behaviour
of the  upper tropospheric layer  (16~km), however, is  noteworthy: it
shows a clear  minimum of turbulence for jet  stream speeds between 20
and 40 m/s, whereas the turbulence increases at both faster and slower
winds. Somehow low  jet stream is `unnatural' for  high atmosphere and
causes more turbulence! Note that  high layer is never perfectly calm,
unlike lower layers.

We also  studied the relation  between ground-layer seeing  and ground
wind as  measured by the  meteo station on  top of Cerro  Tololo. Most
frequently  ground wind blows  from North-East,  otherwise it  is from
South-West;   the  wind-rose  clearly   shows  these   two  dominating
directions. We plot the  ground-layer seeing $\epsilon_g$ against wind
speed     for      these     two     directions      separately     in
Fig.~\ref{fig:gwind}. There is practically no correlation for northern
winds  and  some correlation  for  southern  winds.  Southern wind  is
definitely better for ground-layer seeing.

\section{Summary and conclusions}

For a period of few months we followed with amazement the evolution of
optical turbulence over Cerro Tololo, for the first time being able to
know where the `seeing'  comes from and why it  changes.  The database of
some  28000 low-resolution  TPs,  most of  which  are complemented  by
seeing data, is unique by its volume and time coverage. The insights
gained from these data can be summarized as follows:

\begin{enumerate}
\item
Ground layer turbulence (first 500~m)  at CTIO contributes 60\% of the
total  turbulence  integral  in   50\%  of  cases.  Thus,  a  complete
compensation  of this  layer  would typically  improve  the seeing  
$0.4^{-3/5} =1.7$ times.

\item
The  median  free-atmosphere  seeing  $\epsilon_f$ (all  layers  above
500~m) is $0\farcs 55$,  in 10\% of cases it is  better than $0\farcs 28$, but
it  is practically never  better than  $0\farcs 15$.  The  effective lower
limit to  $\epsilon_f$ is related to the  ever-present weak turbulence
in the upper tropospheric layers above 12~km.

\item
The periods  of stably calm upper  atmosphere with $\epsilon_f<0\farcs
25$ can be as long as few days.  This occurs when the wind velocity at
12~km a.s.l. is around 20-30  m/s.  During these periods, a resolution
gain from ground layer AO compensation will be 2-3.

\item
The  characteristic  time   of  turbulence  variation  increases  with
increasing layer altitudes, from 15 min. (at 50\% correlation level) at
1~km  to 1-2 hours  at 16~km.   Often the  turbulence at  altitudes of
1-8~km has a character of  recurrent strong bursts that last for $\sim
0.5$ hour and repeat every 1-2 hours.

\end{enumerate}

Perhaps the most  important impression from the data  is the fragility
of  astronomical  seeing.  Most  of  the  seeing  results from  local
orographic effects  and is  significantly influenced by  very unstable
ground-layer  turbulence. A  common opinion  that all  good  sites are
similar and  have a median  seeing around $0\farcs 7$ is  in contradiction
with  the  complexity  of   turbulence  phenomena  evidenced  by  this
study. We believe that a  better understanding and modeling of optical
turbulence is possible and will  help to choose `lucky summits' that
are  much  less  affected  by turbulence  generated  near  surrounding
mountains. Statistical data on TP will be essential for this
work.

\section*{Acknowledgments}
 This  work extensively  uses the  data  from CTIO  DIMM developed  by
M.~Boccas, N.~Long, E.~Bustos, and H.E.M.~Schwarz. The support of this
campaign by  the CTIO  telescope operations staff  was crucial  and is
acknowledged.  Development of  MASS  and the  observing campaign  were
supported by the `Sites' program of the New Initiatives Office of NOAO
and managed by A.~Walker.

\bsp

\label{lastpage}

\end{document}